\documentclass[aps,
prd,
preprint,
eqsecnum,
amsmath,
amssymb
]{revtex4}

\usepackage{color}

\usepackage{hyperref}

\begin{document}

\title{Kaluza Ansatz applied to Eddington inspired Born-Infeld
  Gravity} 

\author{Karan Fernandes}
 \email{karan12t@bose.res.in}
\author{Amitabha Lahiri}
\email{amitabha@bose.res.in}
\affiliation{
S. N. Bose National Centre for Basic Sciences,\\
Block-JD, Sector III, Salt Lake, Kolkata-700098, INDIA.
}

\date{\today}

\begin{abstract}
  We apply Kaluza's procedure to Eddington-inspired Born-Infeld
  action in gravity in five dimensions. The resulting action
  contains, in addition to the usual four-dimensional actions for
  gravity and electromagnetism, nonlinear couplings between the
  electromagnetic field strength and curvature.  Considering the
  spherically symmetric solution as an example we find the lowest
  order corrections for the Reissner-Nordstr\"om metric and the
  electromagnetic field.

\end{abstract}

\maketitle

\section{ Introduction}\label{intro}
Eddington-Born-Infeld gravity arose out of a desire to find a
gravitational analog of the determinantal action for
electromagnetism proposed by Born and Infeld~\cite{Born:1934gh},
with the hope that such an action would tame the singularities
arising in gravity in much the same way as the Born-Infeld action
does for electromagnetism. Early approaches in this
area~\cite{Feigenbaum:1997pf, Deser:1998rj, Feigenbaum:1998wy}
proposed a determinantal Lagrangian, of the same form as the
Born-Infeld Lagrangian, but with the electromagnetic tensor being
replaced by the curvature tensor. In particular, models with the
general structure suggested in~\cite{Deser:1998rj} has been
investigated over the years for its cosmological
implications~\cite{Comelli:2004qr}, has been shown to indeed
alleviate the initial cosmological singularity that arises in
standard General Relativity~\cite{Comelli:2004qr, Banados:2010ix,
  Scargill:2012kg}, and has been shown to allow the regulation of
the Schwarzschild singularity for positive
energies~\cite{Wohlfarth:2003ss}. In taking this theory to be not
purely metric, but rather metric-affine~\cite{Banados:2010ix}, it
has been suggested that it has novel implications in the matter
coupling paradigm~\cite{Delsate:2012ky, Delsate:2013bt}.  However,
as has been demonstrated~\cite{Pani:2012qd}, if the theory is taken
to be metric-affine, it still leads to an effective metric theory
upon further expansion. As such, we do not believe that the problem
of coupling matter to gravity in this theory has been resolved or
even adequately addressed. It is to address this issue that we have
undertaken the present work.

We will work with Eddington-inspired Born-Infeld theory, with an
action similar to that of~\cite{Banados:2010ix}, but in five
dimensions. This is then reduced to four dimensions \`a la Kaluza
by compactifying one dimension on a circle. We find corrections to
the four-dimensional Eddington-Born-Infeld theory, highly nonlinear
terms which can be written in the form of infinite sums.

In Sec.~\ref{EBI} we provide a brief review of the
Eddington-Born-Infeld Lagrangian and its equations of motion. We
compare the purely affine Eddington action and the metric-affine
action of Born and Infeld, written in the form
of~\cite{Banados:2010ix}. In the metric-affine theory, the equation
of motion allows the affinity to be written as a function of the
metric, so finally we have an equation for the metric only.  It
turns out that the equations of motion obtained from the two
theories are equivalent, at least in regions of low curvature.  In
Sec.~\ref{kk}, we go over the Kaluza procedure and use it to reduce
a five-dimensional Eddington-inspired Born-Infeld theory to four
dimensions. In Sec.~\ref{eom}, we derive the four-dimensional
equations of motion due to this action. We find deviations from the
gravitational equations as well as from the equations of motion for
the electromagnetic field $A_\mu$ compared to the case when the
electromagnetic field action is simply added to the gravitational
action.

\section{The Eddington Born-Infeld Action}\label{EBI}
Faced with the problem of quantizing the electromagnetic field,
while at the same time ensuring that the theory remain non-singular
at short distances, Born and Infeld~\cite{Born:1934gh} introduced
the action
\begin{equation}
S_{BI} =  \int d^4x \,  b^2 \left[ \sqrt{- \det(g_{\mu
    \nu})} - \sqrt{- \det\left(g_{\mu \nu} + b^{-1}F_{\mu
      \nu}\right)}\,\right]\,,
\label{ebi.BI}
\end{equation}
where $g_{\mu \nu}$ is the metric tensor in a flat spacetime,
$F_{\mu \nu}$ is the electromagnetic field strength tensor, and $b$ 
is a constant which ensures that higher order terms of $F_{\mu
  \nu}$ get smoothed out in the expansion of the square root of the
determinant. This theory, while being a nonlinear generalization of
Maxwell's, has a number of promising features which ensures its
viability. In particular these include the absence of birefringence
in wave propagation and duality
invariance~\cite{BialynickiBirula:1984tx, Plebanski:1970zz,
  Gibbons:1995cv, Deser:1997gq, Deser:1998wv}.

Since any attempt to quantize gravity faces an insurmountable
problem with divergences, it is tempting to try the Born-Infeld
route of ameliorating classical short-distance singularities. A
determinantal action for gravity had been earlier proposed by
Eddington~\cite{Eddington:1924}\,,
\begin{equation}
S_{Edd}= \int d^4x \sqrt{\left| \det
    (R(\Gamma)_{\mu\nu}) \right|}\,.
\label{ebi.Edd}
\end{equation}
Here $ {R(\Gamma)_{\mu\nu}}$ is the symmetric part of the Ricci
tensor constructed as a function of a torsionless affine connection
${\Gamma^{\alpha}_{\beta\gamma}}$\,, but in this action the
${\Gamma}$'s are treated as independent fields, and not as
functions of the metric and its derivatives. Since
Eq. (\ref{ebi.Edd}) is purely affine, we will denote
$R(\Gamma)_{\mu\nu}$ as simply $R_{\mu\nu}$. The Ricci tensor is in
general non-symmetric, so we have to specify that we take its
symmetric part, with inverse defined via
\begin{equation}
 R^{\nu\mu}R_{\mu \alpha} := {\delta}^{\nu}_{\alpha}\,.
\end{equation}

Varying $\Gamma$, we find the equation of motion, 
\begin{equation}
\nabla_{\alpha}\left(\sqrt{\left| \textbf{R} \right|}
R^{\mu\nu}\right) - \nabla_{\beta}\left(\sqrt{\left|
    \textbf{R} \right|}
R^{\beta(\mu}\delta^{\nu)}_{\alpha}\right)=0\,.
\label{ebi.eom}
\end{equation}
Here and in what follows, we have used boldfaced letters to
indicate matrices, and $|\bf{A} |$ to mean the absolute value of
the determinant of the matrix {\bf A}.  We will adopt the matrix
notation wherever convenient and when no confusion can arise, as in
Eq.~(\ref{ebi.eom}) above, where we have denoted the matrix of
$R_{\mu\nu}$ by $\textbf{R}$\,, and the determinant of the said
matrix by $\left| \textbf{R} \right|$\,.  We will write
$\textbf{g}$ when we mean the matrix of $g_{\mu\nu}$\,, but
$\det\textbf{g}$ will be written as $\left| g \right|\,$ in
accordance with common practice.  The second term in
Eq.~(\ref{ebi.eom}) vanishes identically, as can be seen by tracing
over either $\alpha$ and $\mu$, or $\alpha$ and $\nu$. Thus we are
left with the following equation of motion for the connection,
\begin{equation}
\nabla_{\alpha}\left(\sqrt{\left|\textbf{R} \right|}
R^{\mu\nu}\right) = 0\,.
\end{equation}
This equation shows that $\nabla_\alpha$ is the connection for the
`metric' $R_{\mu\nu}$\,, and we may define the metric by a
rescaling
\begin{equation}
R_{\mu\nu} = \lambda g_{\mu\nu}\,.
\end{equation}
Thus the action of Eq.~(\ref{ebi.Edd}) has Einstein spaces as its
extremal points.  The equation of motion is the same as what we get
from the more familiar Einstein-Hilbert action with a cosmological
constant in vacuum,
\begin{equation}
S_{EH} = \frac{1}{16 \, \pi} \int d^4 x \sqrt{\left|
    g \right|}( R - 2 {\Lambda})\,,
\label{ebi.EH}
\end{equation}
provided we set $\lambda = \Lambda\,.$ Here and below, we 
choose units in which $G = 1 \,.$

Eddington's theory thus reproduces Einstein's
equation with a cosmological constant, but only in the absence of
matter. One way of including matter is to generalize the action
in the manner of Deser and Gibbons~\cite{Deser:1998rj}, 
\begin{equation}
S_{DG}= \int d^4x \sqrt{\left| \det \left(g_{\mu\nu} +
    R_{\mu\nu} + X_{\mu\nu}\right) \right|}\,.
\label{ebi.dg}
\end{equation} 
$X_{\mu \nu}$ contains terms quadratic or higher in the curvature,
a `fudge tensor' introduced by hand in order to cancel out the
quadratic curvature terms that arise out of expanding the
determinant, and hence to render the theory ghost free. Matter can
then be added to the theory via a contribution to $X_{\mu\nu}$ from
the matter fields, e.g. $b F_{\mu\nu}$ for the Maxwell field.

A different approach was taken by Banados and Ferreira
\cite{Banados:2010ix}, who introduced, based on earlier
investigations \cite{Vollick:2003qp, Vollick:2005gc,
  Banados:2008fj}, what is now known as the
Eddington-inspired Born-Infeld action,
\begin{equation}
S_{EiBI}=  \frac{1}{8 \pi \kappa} \int d^4x\, \left[\sqrt{\left|
  \textbf{g} + {\kappa} \textbf{R}(\Gamma)\right|}-
\lambda\sqrt{\left| g \right|}\right] \,,
\label{ebi.ebi}
\end{equation} 
where $\kappa > 0$ is a dimensionful constant.
Here again $R_{\mu\nu}$ is a function of an
independent connection ${\Gamma^{\alpha}_{\mu \nu}}$, and thus now
there are two equations of motion --- one from varying with respect
to $g_{\mu\nu}$, and one from varying with respect to the
connection ${\Gamma^{\alpha}_{\mu \nu}}$.  Let us consider the case
where the matter action is added as a separate term,
\begin{equation}
S = S_{EiBI}\left[g_{\mu\nu},\Gamma \right]  +
S_M\left[g_{\mu\nu},\Gamma,\Psi\right]\,.
\label{ebi.ebi2}
\end{equation} 
In the general formalism, the matter action $S_M$ need not be
standard or minimal, but can also depend on the independent
connection $\Gamma$. When $\kappa R_{\mu \nu} \gg g_{\mu \nu}\,,$
the action of Eq.~(\ref{ebi.ebi})
becomes proportional to Eq.~(\ref{ebi.Edd}), and we get Einstein's
equations with cosmological constant $\Lambda$ by setting ${\kappa}
= \frac{16 \pi}{\Lambda}$.  In finding this limit, we simply
consider $\frac{\lambda}{\kappa} \sqrt{\left| g \right|}$ to be
negligible in comparison with $\kappa \sqrt{\left| \textbf{R}
  \right|}$.

On the other hand, when $\kappa R_{\mu \nu} \ll g_{\mu \nu}\,,$ we
can expand the determinant in a power series. Then $\kappa$ counts
the power of curvature appearing in each term of the expansion.  At
the lowest order we find the Einstein-Hilbert action
Eq.~(\ref{ebi.EH}), as we should, provided we set $\lambda=
\kappa\Lambda +1$. Since in this paper we are concerned with
corrections to Einstein gravity stemming from Eq.~(\ref{ebi.ebi}),
we will fix $\lambda = \kappa \Lambda +1$ in what follows. To 
understand the difficulties of adding matter to this theory, let us
follow~\cite{Banados:2010ix} for the moment and take $S_M$ in
Eq.~(\ref{ebi.ebi2}) to be a standard matter action. 

Then the stress energy tensor $T_{\mu \nu}$ is calculated by 
varying $S_M$ with respect to the metric,
\begin{equation}
T_{\mu \nu} := - \frac{2}{\sqrt{\left| g \right|}} \frac{\delta
  S_M}{\delta g^{\mu \nu}} \, . 
\label{ebi.set}
\end{equation}
For numerical factors, we will adopt the conventions of~\cite{Wald},
where the electromagnetic stress energy tensor and the Maxwell
action take the following form
\begin{equation}
S_{EM} = - \frac{1}{16 \pi} \int d^4 x \sqrt{\left| g \right|} 
F_{\alpha \beta} F^{\alpha \beta} \,, \qquad
T_{\mu \nu} = \frac{1}{4 \pi} \left( F_{\mu \alpha} 
F_{\nu}^{\phantom{\nu} \alpha} - \frac{1}{4} 
F_{\alpha \beta} F^{\alpha \beta} g_{\mu \nu} \right) \,.
\label{ebi.Tdef}
\end{equation}
These will be relevant in the sections to follow. For now, 
we will continue the general exposition for any matter 
field, for which only Eq.(\ref{ebi.set}) will be of relevance.


With the assumption of a standard matter action, we obtain the
following equations of motion from Eq.~(\ref{ebi.ebi2}), after
varying with respect to the metric and the connection (both being
independent of each other at this point)
\begin{align}
\sqrt{\left| \textbf{g} + {\kappa}
    \textbf{R}(\Gamma)\right|}\left((\textbf{g} +
{\kappa}\textbf{R}(\Gamma))^{-1}\right)^{\mu\nu} - (\kappa \Lambda 
+1)\sqrt{\left| g \right|} g^{\mu\nu} &=
- 8 \pi \kappa \sqrt{\left| g \right|} T^{\mu\nu} \,,
\label{ebi.var1}\\
\nabla_{\alpha}(\sqrt{\left| \textbf{g} + {\kappa}
    \textbf{R}(\Gamma)\right|}\left((\textbf{g} +
{\kappa}\textbf{R}(\Gamma))^{-1}\right)^{\mu\nu}) &= 0\,.
\label{ebi.var2}
\end{align}
As explained earlier, the boldfaced letters symbolize the
corresponding matrices. 

Since the matter action is independent of the connection
$\Gamma$\,, it is possible to solve for the connection in the same
way as was done in the Eddington case. We take
$q_{\mu\nu}=g_{\mu\nu} + \kappa R(\Gamma)_{\mu\nu}$, and require
that it satisfy Eq.~(\ref{ebi.var2}). This gives a connection
\begin{equation}
\Gamma^{\alpha}_{\beta\gamma} = \frac {1}{2}q^{\alpha\mu}
[q_{\beta\mu,\gamma} + q_{\gamma\mu,\beta} - q_{\beta\gamma,\mu}]\,,
\label{ebi.conn}
\end{equation}
and Eq.~(\ref{ebi.var1}) takes the form
\begin{equation}
\sqrt{\left| \textbf{q} \right|} q^{\mu\nu} =
(\kappa \Lambda +1) \sqrt{\left| g \right|}
g^{\mu\nu}  - 8 \pi \kappa\sqrt{\left| g \right|}
T^{\mu\nu}\,,
\label{ebi.qdef}
\end{equation}
where $q^{\mu \nu} = ((\textbf{g} +
{\kappa}\textbf{R}(\Gamma))^{-1})^{\mu\nu}$.  The left hand side of
this equation depends on both the metric and the independent
connection, since the auxilliary metric $q_{\mu\nu}$ introduced
here is a function of both $g_{\mu\nu}$ and $R(\Gamma)_{\mu\nu}$,
whereas the right hand side depends only on the metric. This
suggests that the connection is not truly independent of the
metric; and this is indeed the case, as was shown in
~\cite{Banados:2010ix, Pani:2012qd}. First we find the determinant
of Eq.~(\ref{ebi.qdef}), for which we get the following expression
\begin{equation}
\left| \textbf{q} \right| = \left|  g \right|^2 
(\left| (1 + \kappa \Lambda) {\textbf{g}}^{-1} - 8 \pi \kappa ~
  {\textbf{g}}^{-1}\textbf{T}{\textbf{g}}^{-1} \right|)\,. 
\label{ebi.qdet}
\end{equation}

Substituting this in Eq.~(\ref{ebi.qdef}), we find the following
expression for $q_{\mu \nu}$\,,
\begin{equation}
 \textbf{q} =  \sqrt{\left| g \right|}
  \sqrt{\left| (1 + \kappa \Lambda) \textbf{g}^{-1}- 8 \pi \kappa
     \textbf{g}^{-1} \textbf{T} \textbf{g}^{-1} \right|} {\left(
     (\kappa 
 \Lambda +1) \textbf{g}^{-1}  - 8 \pi \kappa 
  \textbf{g}^{-1}\textbf{T}\textbf{g}^{-1}\right)}^{-1}\,.
\label{ebi.qform}
\end{equation}
We can now expand this result by using the standard
formul\ae\ for the square root of the determinant, $(\left|
  \textbf{I + A} \right|)^{\frac{1}{2}} = 1 +
\frac{\text{tr(\textbf{A})}}{2} + \frac{\text{(tr
    (\textbf{A}))}^2}{8} - \frac{\text{tr}(\textbf{A}^2)}{4} +
\mathcal{O}({\textbf{A}}^3) $ , and the inverse of a sum of
matrices, $(\textbf{A} + \textbf{B})^{\mu\nu} = A^{\mu \nu} -
A^{\mu \alpha}B_{\alpha \beta}A^{\beta \nu} + A^{\mu
  \alpha}B_{\alpha \beta}A^{\beta \gamma} B_{\gamma\delta}A^{\delta
  \nu} + \mathcal{O}({\textbf{B}}^3)$\,, where \textbf{I} is the 
identity matrix. Using these, we acquire the expression for
$R(\Gamma)_{\mu\nu}$ as
\begin{equation}
R(\Gamma)_{\mu\nu} = \Lambda g_{\mu\nu} +  8\pi \left[ T_{\mu\nu} 
- \tfrac{1}{2} T g_{\mu\nu} \right] + 64 {\pi}^2 \kappa \left[
S_{\mu\nu}  
- \tfrac{1}{4} S g_{\mu\nu}\right]
+ \mathcal{O}({\kappa}^2)\,,
\label{ebi.ric2}
\end{equation}
where $S_{\mu\nu}$ is given by
\begin{equation}
S_{\mu\nu} = T_{\mu\alpha}T^{\alpha}_{\phantom{\alpha}\nu} -
\frac{1}{2} T T_{\mu\nu}\,.
\end{equation}
However, Eq.~(\ref{ebi.ric2}) can also be used to find the
expression for the Ricci tensor as a function of the
metric. Inverting Eq.~(\ref{ebi.qform}) gives us the expression for
$q^{\mu\nu}$. Keeping only terms to order $\kappa\,,$ we find
\begin{equation}
q^{\mu \nu} = g^{\mu \nu} - \kappa {\tau}^{\mu \nu} +
\mathcal{O}({\kappa}^2) 
\end{equation}
where $\tau_{\mu\nu} = \Lambda g_{\mu\nu} + 8 \pi \left[ T_{\mu\nu}
  - \frac{1}{2} Tg_{\mu\nu} \right]$. One can now use the
expressions for $q_{\mu\nu}$ and $q^{\mu\nu}$ in
Eq.~(\ref{ebi.conn}), and expand up to order $\kappa$\,.  After a
bit of algebra, this produces the expression
\begin{equation}
\Gamma^\alpha_{\beta\gamma} = 
\left\{\buildrel{\alpha}\over{_{\beta\gamma}}\right\}
+ \frac12 \kappa q^{\alpha\delta}
\left(R(\Gamma)_{\delta\beta;\gamma}
+ R(\Gamma)_{\gamma\delta;\beta}
- R(\Gamma)_{\beta\gamma;\delta}\right)\,,
\end{equation}
where the semicolon in the subscript implies a covariant derivative
calculated using
$\left\{\buildrel{\alpha}\over{_{\beta\gamma}}\right\}\,,$ the
Christoffel symbols corresponding to $g_{\mu\nu}\,.$ The Ricci
tensor calculated using these $\Gamma$ is given by
\begin{equation}
R(\Gamma)_{\mu\nu} = R(g)_{\mu\nu} + \frac12 \kappa
g^{\alpha\beta} \left(R(\Gamma)_{\alpha\mu;\nu\beta} 
+ R(\Gamma)_{\alpha\nu;\mu\beta} 
  -R(\Gamma)_{\mu\nu;\alpha\beta} -R(\Gamma)_{\alpha\beta;\mu\nu} 
\right) +  {\cal{O}}(\kappa^2)\,.  
\label{ebi.ric3}
\end{equation}
Equating the right hand sides of Eq.~(\ref{ebi.ric2}) and
Eq.~(\ref{ebi.ric3}) we find
\begin{align}
  R(g)_{\mu\nu} &= \Lambda g_{\mu\nu} +  8 \pi \left[ T_{\mu\nu} -
    \tfrac {1}{2} T g_{\mu\nu} \right] + 64 {\pi}^2 \kappa 
\left[S_{\mu\nu} - \tfrac {1}{4} S 
    g_{\mu\nu}\right] \nonumber \\
& \qquad \qquad + \tfrac{1}{2} \kappa
  \left[\nabla_{\mu}\nabla_{\nu}\tau - 
    2 \nabla^{\alpha}\nabla_{(\mu}\tau_{\nu)\alpha} +
    \Box\tau_{\mu\nu}\right] + \mathcal{O}({\kappa}^2)\,,
\label{ebi.ricq}
\end{align}
where we have written $\Box \equiv \nabla_\mu\nabla^\mu\,,$ and
$\tau_{\mu\nu}$ is as defined above.  We see that this expression
contains at least third derivatives of the matter fields. A
consequence of this is that there could exist singularities in the
curvature invariants should the matter distribution be
discontinuous enough, and that there are surface singularities, in
the case of polytropic stars~\cite{Pani:2012qd}.  This has brought
the viability of this theory into question.

We note here that there is another way of writing the equation of
motion, which follows from the fact that Eq.~(\ref{ebi.ric3})
implies that to leading order, $R(\Gamma)_{\mu\nu} =
R(g)_{\mu\nu}\,,$ and all corrections to this are ${\cal
  O}(\kappa)$ or higher. Thus we can rewrite Eq.~(\ref{ebi.ric3})
as 
\begin{equation}
R(\Gamma)_{\mu\nu} = R(g)_{\mu\nu} + \frac12 \kappa
g^{\alpha\beta} \left(R(g)_{\alpha\mu;\nu\beta} 
+ R(g)_{\alpha\nu;\mu\beta} 
  -R(g)_{\mu\nu;\alpha\beta} -R(g)_{\alpha\beta;\mu\nu} 
\right) +  {\cal{O}}(\kappa^2)\,.  
\label{ebi.ric4}
\end{equation}
Putting this back into Eq.~(\ref{ebi.ric2}), we can write
the equation of motion as
\begin{align}
R(g)_{\mu\nu} &= \Lambda g_{\mu\nu} +  8\pi \left[ T_{\mu\nu} 
- \tfrac{1}{2} T g_{\mu\nu} \right] + 64 {\pi}^2 \kappa \left[
S_{\mu\nu}  
- \tfrac{1}{4} S g_{\mu\nu}\right] \notag \\
&\qquad 
- \frac12 \kappa
g^{\alpha\beta} \left(R(g)_{\alpha\mu;\nu\beta} 
+ R(g)_{\alpha\nu;\mu\beta} 
  -R(g)_{\mu\nu;\alpha\beta} -R(g)_{\alpha\beta;\mu\nu} 
\right) +  {\cal{O}}(\kappa^2)\,.  
\label{ebi.eomfinal}
\end{align}
Any solution of Eq.~(\ref{ebi.ricq})
is a solution of Eq.~(\ref{ebi.eomfinal}) and vice versa. Although
both these equations have been derived from Eq.~(\ref{ebi.ric2}) by
neglecting ${\cal O}(\kappa^2)$ terms, it is obvious that we can
follow the procedure to get two equations at higher order in
$\kappa$ as well. While the $\kappa$ expansion in the first
equation corresponds to a series in the stress-energy tensor and
its derivatives, whereas the expansion in the second equation is
one in curvature. 

Since Eq.~(\ref{ebi.eomfinal}) is written fully in terms of the
Levi-Civita connection, we can use the relation between this
connection and $R(g)_{\mu\nu}$ to rewrite it as
\begin{align}
R_{\mu\nu} = & \Lambda g_{\mu\nu} + 8 \pi \left(T_{\mu\nu} -
  \frac{1}{2}T 
g_{\mu\nu}\right) + 64 {\pi}^2 \kappa \left(S_{\mu\nu} -
\frac{1}{4}S g_{\mu\nu} \right)  
\nonumber\\ 
 & \qquad \qquad + \frac{1}{2}\kappa 
\left[2 R_{\alpha \mu \beta \nu}   
R^{\alpha \beta} -  2R_{\mu \beta} R^{\beta}_{\nu} 
 + \Box R_{\mu \nu}\right] +
 \mathcal{O}({\kappa}^2) \,.
\label{ebi.eomfinal2}
\end{align}

We have seen above that if we start from the metric-affine theory,
the equations of motion naturally lead to a purely metric
expression for the usual Ricci tensor. It is thus natural to
investigate what the equations of motion might be if we started
with a purely metric version of the theory. The action is the same
as in Eq.~(\ref{ebi.ebi2}), but with $R_{\mu\nu} =
R(g)_{\mu\nu}$\,,
\begin{equation}
S = \frac{1}{8 \pi \kappa} \int d^4x\, \left[\sqrt{\left|
  \textbf{g} + {\kappa} \textbf{R}(g)\right|}-
\lambda\sqrt{\left| g \right|}\right] +
S_M\left[g_{\mu\nu},\Psi\right]\,.
\label{ebi.ebi2g}
\end{equation} 
As before, we set $\lambda = \kappa \Lambda +1$\,, and vary this
action with respect to $g_{\mu\nu}\,,$ to find
\begin{align}
\delta S &= 
\frac{1}{8 \pi \kappa}\int d^4x \left[
\frac{1}{2}\sqrt{\left| \textbf{g} + {\kappa}
      \textbf{R}\right|}(\textbf{g} +
{\kappa}\textbf{R})^{-1})^{\mu\nu} (\delta g_{\mu\nu} + 
\kappa\,\delta R_{\mu\nu})\right. \nonumber\\
 & \qquad\qquad\qquad\qquad \left. - \frac{1}{2}
(\kappa \Lambda  +1)\sqrt{\left| g \right|} g^{\mu\nu} \delta
g_{\mu\nu} + \frac{8 \pi \kappa }{2} \sqrt{ \left|
      g \right|}T^{\mu\nu} \delta g_{\mu\nu}\right] 
\nonumber\\
&= \frac{1}{16 \pi \kappa} \int d^4x \left[\sqrt{\left|
      \bar{\textbf{q}}\right|}{\bar{q}}^{\mu\nu} + \kappa
  H^{\mu\nu} - (\kappa \Lambda +1) \sqrt{\left| g \right|}
  g^{\mu\nu} + 8 \pi \kappa \sqrt{\left| g \right|}
  T^{\mu\nu}\right] 
\delta g_{\mu\nu}\,,
\label{ebi.dsebi}
\end{align}
where we have defined ${\bar{q}}_{\mu\nu} = g_{\mu\nu} + \kappa
R_{\mu\nu}$, to distinguish it from the earlier case where we had
$R(\Gamma)_{\mu\nu}$. We have also defined $H^{\mu\nu}$, which in
terms of the notation just introduced, is given by
\begin{equation}
  H^{\mu\nu} = \frac{1}{2} [\nabla_{\alpha} \nabla^{\mu}
  (\sqrt{\left|\bar{\textbf{q}}\right|}\bar{q}^{\alpha\nu}) +
  \nabla_{\alpha} 
  \nabla^{\nu}(\sqrt{\left|\bar{\textbf{q}}\right|}
 \bar{q}^{\mu\alpha})  - \nabla_{\alpha}\nabla_{\beta}
  (\sqrt{\left|\bar{\textbf{q}}\right|}\bar{q}^{\beta \alpha}
  g^{\mu\nu}) - \Box 
  (\sqrt{\left|\bar{\textbf{q}}\right|}\bar{q}^{\mu\nu})]\,.
\label{ebi.hdef}
\end{equation}
In going from the first line to the second line of
Eq.~(\ref{ebi.dsebi}), we made use of the Palatini identity $\delta 
R_{\mu\nu} = \nabla_{\alpha}(\delta \Gamma_{\mu \nu}^{\alpha}) -
\nabla_{\mu}(\delta \Gamma_{\alpha \nu} ^{\alpha})$, and exploited
the Leibniz rule for covariant derivatives to eliminate total
derivatives.

We will now make use of the equation of motion that comes out of 
this, 
\begin{equation}
\sqrt{\left|\bar{\textbf{q}}\right|}\bar{q}^{\mu\nu} + 
\kappa H^{\mu\nu} - (\kappa \Lambda +1) 
\sqrt{\left| g \right|} g^{\mu\nu} = -
8 \pi \kappa \sqrt{\left| g \right|} T^{\mu\nu}\,,
\label{ebi.ebieom}
\end{equation}
to find an expression for $R_{\mu\nu}$. We can substitute
${\bar{q}}_{\mu\nu}$ in the expression for $H^{\mu\nu}$ above to
find
\begin{align}
H^{\mu\nu} &= -\frac{1}{2}\kappa \sqrt{\left|
      g \right|}\left[ 2 
R^{\mu\phantom{\alpha\delta}\nu}_{\phantom{\mu}\alpha\delta}
R^{\alpha \delta} + 2R^{\mu}_{\delta}R^{\nu \delta} - 
\Box R^{\mu\nu} + \frac{1}{2} g^{\mu\nu}\Box R\right]
+\mathcal{O}({\kappa}^2) \nonumber\\ 
&\equiv \sqrt{\left|
      g \right|} \tilde{H}^{\mu\nu}\,.
\label{ebi.hexp}
\end{align}
Eq.~(\ref{ebi.hexp}) clearly shows that there are no
$\mathcal{O}({\kappa}^0)$ terms in $H^{\mu\nu}$ in the lowest
order. Eq.~(\ref{ebi.ebieom}) will thus also yield Einstein's
equation with a cosmological constant in the case of $\kappa =
0$. To make things more explicit, we proceed as before and acquire
the determinant of Eq.~(\ref{ebi.ebieom}) 
\begin{equation}
\bar{\textbf{q}} =  \sqrt{\left| g \right|}
  \sqrt{\left| (1 + \kappa \Lambda) \textbf{g}^{-1}- \kappa
     \textbf{g}^{-1}(\tilde{\textbf{H}} + 8 \pi \, \textbf{T})
     \textbf{g}^{-1} \right|}{\left( (\kappa \Lambda +1) \textbf{g}^{-1}
 - \kappa \textbf{g}^{-1}(\tilde{\textbf{H}} + 8 \pi \,
  \textbf{T})\textbf{g}^{-1} \right)}^{-1}\,. 
\label{ebi.qform2}
\end{equation}
The expansion of  Eq.(\ref{ebi.qform2}) up to order $\kappa$
reveals the following expression 
\begin{align}
R_{\mu\nu} = & \Lambda g_{\mu\nu} + 8 \pi \left(T_{\mu\nu} -
  \frac{1}{2}T 
g_{\mu\nu}\right) + 64 {\pi}^2 \kappa \left(S_{\mu\nu} -
\frac{1}{4}S g_{\mu\nu} \right)  
\nonumber\\ 
 & \qquad \qquad + \frac{1}{2}\kappa 
\left[2 R_{\alpha \mu \beta \nu}   
R^{\alpha \beta} -  2R_{\mu \beta} R^{\beta}_{\nu} 
 + \Box R_{\mu \nu}\right] +
 \mathcal{O}({\kappa}^2) \,.
\label{ebi.Rexp}
\end{align}
This is the same equation that we found in the metric-affine theory
when we wrote the equation of motion in terms of quantities derived
from $g_{\mu\nu}\,.$ Thus the metric theory and the metric-affine
theory are equivalent. 

This observation brings us to the main motivation for this work. We
ask if there is a natural way of incorporating the matter part of
the action into the theory other than simply adding it, such that
we still reproduce Einstein's theory in the weak limit. There exist
still further proposals over the incorporation of matter in this
theory.  In~\cite{Vollick:2003qp, Vollick:2005gc},
$R(\Gamma)_{\mu\nu}$ was allowed to have an antisymmetric
component, leading to the action for a massive vector field. In a
different approach~\cite{Delsate:2012ky, Delsate:2013bt}, matter
was coupled to the ``metric" $q_{\mu\nu}$ in the field equations
Eq.~(\ref{ebi.var1}) and Eq.~(\ref{ebi.var2}).  Since the vacuum
equations are the same as in usual general relativity, this
coupling plays out only in signficantly matter dense regions, as in
the interior of stars.

Here we take a geometric approach, while staying close to the
original Born-Infeld idea of ameliorating singularities. Our
approach will be to use Kaluza's idea~\cite{Kaluza:1921tu} of
unifying gravity and electromagnetism in a five-dimensional theory
of gravitation, and apply it to the five-dimensional
Eddington-Born-Infeld theory. Since this procedure deals only with
the five-dimensional metric, we will necessarily deal with the
metric version of the Eddington-Born-Infeld action. However, as we
have seen in this section, the two approaches agree at least to
${\cal O}(\kappa)\,,$ so we may consider the resulting action a
natural way of incorporating electromagnetic fields in the
four-dimensional Eddington-inspired Born-Infeld gravitational
theory.

\section{The Kaluza Ansatz}\label{kk}
We start by writing the metric in the form
\begin{equation}
\hat{g}_{AB} = 
\begin{pmatrix}
g_{\mu\nu} + {\alpha}^2 {\Phi}^2 A_{\mu}A_{\nu}  &  {\alpha}
{\Phi}^2 A_{\mu} \\ 
{\alpha} {\Phi}^2 A_{\nu} &  {\Phi}^2 
\end{pmatrix}\,.
\label{kk.ansatz}
\end{equation}
Here and later, uppercase Latin indices are five-dimensional,
$A,B,\dots = 0,\cdots,3,5,$ while Greek indices are
four-dimensional, $\mu,\nu,\dots = 0,\cdots,3$\,.  Five-dimensional
objects will be written with hats, and ${\alpha}$ is a parameter
which will be fixed later. The inverse of this matrix is
\begin{equation}
\hat{g}^{AB} = 
\begin{pmatrix}
g^{\mu\nu}  & - {\alpha} A^{\mu} \\
- {\alpha} A^{\nu} &  {\alpha}^2 A^{\gamma} A_{\gamma} +
\frac{1}{{\Phi}^2} 
\end{pmatrix}\,.
\label{kk.ginv}
\end{equation}
While the consequences of the including the scale of the fifth 
dimension as an independent scalar field~\cite{Jor} is
interesting in its own right, our interest lies in the coupling of
electromagnetism to gravity, so we will set ${\Phi}$ = 1.  In
Appendix~\ref{appb} we have given the expression for the Ricci 
scalar for a non-trivial $\Phi$\,. We will construct the 
Eddington-Born-Infeld action for the five-dimensional metric 
theory, i.e. we will write Eq.~(\ref{ebi.dsebi}) for the above
 metric ansatz and derive some of its consequences.

The Ricci tensor components are calculated in a straightforward
manner, 
\begin{align}
\hat{R}_{\mu\nu} &= R_{\mu\nu} + \frac{1}{4}{\alpha}^4
  F^{\beta\gamma}F_{\beta\gamma} A_{\mu}A_{\nu}
-\frac{1}{2}{\alpha}^2 (A_{\mu}{\nabla}_{\beta}
F^{\beta}_{\phantom{\beta}\nu} + A_{\nu}{\nabla}_{\beta}
F^{\beta}_{\phantom{\beta}\mu} +
F_{\beta\mu}F^{\beta}_{\phantom{\beta} \nu}) \notag \\
\hat{R}_{\mu 5} &= \frac{1}{4}{\alpha}^3 F^{\beta \gamma}F_{\beta
  \gamma} A_{\mu} - \frac{1}{2}{\alpha}
({\nabla}_{\beta}F^{\beta}_{\phantom{\beta}\mu})\,, \qquad
\hat{R}_{55} = \frac{1}{4}{\alpha}^2 F^{\beta \gamma} F_{\beta
  \gamma}\,,
\end{align}
giving the Ricci scalar,
\begin{equation}
\hat{R} = R - \frac{{\alpha}^2}{4} F^{\beta \gamma} F_{\beta
  \gamma}\,.
\end{equation}
If we write the radius of compactification of the fifth dimension
as $\tilde R\,,$ and the five-dimensional Newton's constant as
$\hat G_5\,,$ we find that setting 
\begin{equation}
 \frac{2 \pi \tilde{R}}{{\hat{G}}_{5}} = \frac{1}{G} = 1 \: ,
 \qquad {\alpha}^2 = 4G = 4 
\label{kk.G5}
\end{equation}
leads to the reduction of the five-dimensional Einstein-Hilbert
action as 
\begin{align}
\frac{1}{16 \pi {\hat{G}}_{5}} \int d^5x
\sqrt{\left| \hat{g} \right|} \hat{R} = 
\frac{1}{16 \pi} \int d^4x \sqrt{\left| g \right|} R -
\frac{1}{16 \pi}\int d^4x \sqrt{\left| g \right|}
 F^{\beta \gamma} F_{\beta \gamma} \,.
\label{kk.eh}
\end{align}
The factor in front of the electromagnetic action agrees with our
conventions, shown in Eq.~(\ref{ebi.Tdef}).  We will set this value
of ${\alpha}^2$ in the remainder of the paper.

We now write the Eddington-inspired Born-Infeld action in the
metric form, i.e. the action of Eq.~(\ref{ebi.ebi2g}) without
$S_M$, but in five dimensions. Then we will need to  find two
determinants -- those of $\hat{g}_{AB}$ and $\hat{q}_{AB} =
\hat{g}_{AB} + {\kappa} \hat{R}_{AB}$.  Using the usual
decomposition of a block matrix, 
\begin{align}
\det\begin{pmatrix}
A  &  B \\
C &  D
\end{pmatrix}
&=
\det\left[\begin{pmatrix}
A - BD^{-1}C  &  BD^{-1} \\
0 & 1
\end{pmatrix}
\begin{pmatrix}
1 & 0\\
C & 1
\end{pmatrix}
\begin{pmatrix}
1  & 0\\
0 &  D
\end{pmatrix}\right] \notag\\
&= \det(D) \det(A-B D^{-1} C)\,,
\end{align}
we find that $ \left| \det(\hat{g}_{AB}) \right| = 
\left|\det(g_{\mu\nu})\right| \equiv \left| g \right|$. 
The other block matrix, $\hat{q}_{AB}$, has the following 
components
\begin{align}
\hat{q}_{\mu\nu} &= g_{\mu\nu} + 4 A_{\mu}A_{\nu}
\left(1+ {\kappa} F^2\right) +
{\kappa} \left[ R_{\mu\nu} -
2 \left(A_{\mu}{\nabla}_{\beta}
F^{\beta}_{\phantom{\beta}\nu}  
+ A_{\nu}{\nabla}_{\beta} F^{\beta}_{\phantom{\beta}\mu} +
F_{\beta \mu}F^{\beta}_{\phantom{\beta}\nu}\right) \right] \notag
\\ 
\hat{q}_{5 \nu} &= 2 A_{\nu}
\left(1+ {\kappa} F^2\right) -
\kappa{\nabla}_{\beta}
F^{\beta}_{\phantom{\beta}\nu} \,, \qquad 
\hat{q}_{55}= 1 +  {\kappa} F^2\,,
\label{kk.q55}
\end{align}
where we have written $F_{\mu\nu}F^{\mu\nu} = F^2\,.$ 
Now we make a formal expansion in powers of $\kappa$ to find 
$(\hat{q}_{55})^{-1}$, and get
\begin{equation} 
  \left[q_{\mu\nu} - q_{\mu 5} (q_{55})^{-1} q_{5 \nu}\right] =
  g_{\mu \nu} + {\kappa}\left(R_{\mu\nu} + 
  2  F_{\mu\beta}F^{\beta}_{\phantom{\beta}\nu}\right) +
  {\nabla}_{\delta}F^{\delta}_{\phantom{\delta}\mu}
  {\nabla}_{\beta}F^{\beta}_{\phantom{\beta}\nu}
  \displaystyle\sum_{n=0}^{\infty} 
  \left(-1\right)^{n + 1} {\kappa}^{n+2} F^{2n}\,. 
\label{kk.qexact}
\end{equation}
Using this, we can write the Eddington-Born-Infeld action in the  
five-dimensional space-time,
\begin{align}
S &= \frac{1}{8 \pi \hat{G}_{5} \kappa} \int d^5x [\sqrt{\left|
    (\hat{g}_{AB} + {\kappa}\hat{R}_{AB}) \right|} - 
({\kappa}{\Lambda} + 1)\sqrt{\left|
    \hat{g} \right|}] \nonumber\\
 &= \frac{1}{8 \pi \kappa} \int d^4x \left[\sqrt{1 +
  \kappa F^2}\times\right. \nonumber \\
& \left.\sqrt{\left| g_{\mu
      \nu} + {\kappa}(R_{\mu \nu} + 2  F_{\mu
      \beta} F^{\beta}_{\phantom{\beta}\nu}) +
    ({\nabla}_{\delta}F^{\delta}_{\phantom{\delta}\mu}
    {\nabla}_{\beta}F^{\beta}_{\phantom{\beta}\nu})
\sum_{n=0}^{\infty} 
    \left(-1 \right)^{n + 1} {\kappa}^{n+2}
    F^{2n}   \right|} -
({\kappa}{\Lambda}+1)\sqrt{g}\right]\,. 
\label{kk.ebiaction}
\end{align}
We have used Eq.~(\ref{kk.G5}) in going from the first to the
second line in Eq.~( \ref{kk.ebiaction}), analogous to the
Einstein-Hilbert treatment above.  Remember that $\kappa$ counts
the powers of curvature, so keeping terms up to some given order of
$\kappa$ is the same as neglecting higher powers of curvature. We
will be interested in determining the lowest order corrections to
the equations of motion, which means that we need only expand to
second order, i.e.  ${\cal O}({\kappa}^2)$\,. In order to get an
${\cal O}({\kappa}^2)$ contribution from the first term, we need
only consider the $n=0$ term in the sum.  The action to this order
is given by
\begin{equation} 
  S=\frac{1}{8 \pi {\kappa}} \int d^4x \left[\sqrt{1 +
     \kappa F^2}\sqrt{\left| g_{\mu \nu} 
+ {\kappa}\left(R_{\mu\nu} +  2 F_{\mu \beta}
        F^{\beta}_{\phantom{\beta}\nu}\right) - {\kappa}^2
               {\nabla}_{\delta}F^{\delta}_{\phantom{\delta}\mu}
        {\nabla}_{\beta}F^{\beta}_{\phantom{\beta}\nu} \right|}
    -  ({\kappa}{\Lambda}+1)\sqrt{\left| g \right|}\right]\,.
\label{kk.action}
\end{equation}
%

\section{Equations of motion}\label{eom}
We will first expand Eq.~(\ref{kk.action}) up to first order in
$\kappa$, before proceeding to its expansion to order ${\kappa}^2$.
In expanding the
determinant and varying the action, we also expect that linearity
will ensure that we get the same equations of motion at the lowest
order as Eq.~(\ref{ebi.ebi}), which is nothing but Einstein's
equation with a cosmological constant.

If we expand all terms to first order $\kappa$, the action is given 
by
\begin{equation}
 S= \frac{1}{8 \pi {\kappa}} \int d^4x  \sqrt{g} \left[\left(1 +
     \frac{\kappa}{2} F^2\right)\left(1 + \frac{\kappa}{2} 
\left(R - 2 F^2\right) \right) - (\kappa \Lambda + 1) 
+\mathcal{O}({\kappa}^2) \right]  \,,
 \end{equation}
from which we get the equations of motion
\begin{equation}
G_{\mu\nu} = -\Lambda g_{\mu\nu} +  8 \pi T_{\mu\nu}\,\, , 
\qquad \nabla_{\alpha}F^{\alpha \nu} = 0 \, .
\label{eom.einmax}
\end{equation}
We thus recover the equations of motion of the Einstein-Maxwell
theory, and dynamics is unaffected at the lowest order, as it
should be.

Expanding to second order in $\kappa$ gives us
\begin{align}
  S &= \frac{1}{8 \pi {\kappa}} \int d^4x \sqrt{g} \,\left[\left(1 +
      \frac{{\kappa} F^2}{2} - \frac{{\kappa}^2 F^4}{8}\right)
    \left(1 + \frac{\kappa}{2} (R - 2 F^2) + \frac{{\kappa}^2}{8}
      (R^2 + 4 F^4 -
      4RF^2)\right.\right. \nonumber \\
  & \qquad \qquad -
  \left.\left.\frac{{\kappa}^2}{4}(R^{\alpha\beta}R_{\beta\alpha} +
      4
      R^{\alpha\beta}F_{\beta\gamma}F^{\gamma}_{\text{\space\space}\alpha}
      + 4 F^{\alpha\beta}F_{\beta\gamma}F^{\gamma\delta}F_{\delta\alpha}
      - 2 \nabla_{\alpha} F^{\alpha \beta}
      \nabla^{\gamma} F_{\gamma \beta})\right)\right. \nonumber \\
  & \quad\qquad\qquad\qquad \left. - (\kappa \Lambda + 1)
    +{\mathcal O}({\kappa}^3) \right] \,,
\label{eom.ksquare}
\end{align}
which is extremized with respect to the inverse metric, to get the
following equation of motion to order ${\kappa}$
\begin{equation}
G_{\mu\nu} = - \Lambda g_{\mu\nu} + 8 \pi T_{\mu\nu} + \kappa
P_{\mu \nu} +\kappa Q_{\mu \nu}\,. 
\label{eom.eom}
\end{equation}
Here $G_{\mu\nu}$ is the usual Einstein tensor, $T_{\mu\nu}$ is
the usual energy-momentum tensor of electrodynamics, and
\begin{align}
P_{\mu\nu} &= R_{\mu\alpha}R_{\nu}^{\alpha} -\tfrac{1}{2} R
R_{\mu\nu} -\tfrac{1}{4} R_{\alpha \beta} R^{\alpha \beta} g_{\mu
  \nu} + \tfrac{1}{8} R^2 g_{\mu \nu} + \tfrac{1}{2} \nabla_{\mu}
\nabla_{\nu} R - \tfrac{1}{2} g_{\mu \nu} \Box R  
\notag \\
&  \qquad - \nabla_{\alpha} \nabla_{(\mu} R^{\alpha}_{\nu)} + \tfrac{1}{2}
\Box R_{\mu \nu} + \tfrac{1}{2} g_{\mu \nu} \nabla_{\alpha}
\nabla_{\beta} R^{\alpha \beta}
\,, \\
Q_{\mu \nu} &=  R F_{\mu \alpha} F_{\nu}^{\phantom{\nu} \alpha}
+ \nabla_{\alpha} F^{\alpha}_{\phantom{\alpha} \nu}
\nabla_{\beta}F^{\beta}_{\phantom{\beta} \mu} + 2
R_{\mu\alpha}F^{\alpha\beta}F_{\beta\nu} + 2
F_{\mu\alpha}R^{\alpha\beta}F_{\beta\nu} \notag \\ 
& \qquad + 2 F_{\mu\alpha}F^{\alpha\beta}R_{\beta\nu} + 8 F_{\mu
  \alpha}F^{\alpha \beta}F_{\beta
  \gamma}F^{\gamma}_{\phantom{\gamma} \nu} - 2 \nabla_{(\mu}(F_{\nu
  ) \beta} \nabla_{\alpha}F^{\alpha \beta}) \notag \\ 
& \qquad - 2 \nabla_{\alpha}(F^{\alpha}_{\phantom{\alpha}(\mu}
\nabla_{|\beta|}F^{\beta}_{\phantom{\beta}\nu)}) + F^2 F_{\mu\beta}
F_{\nu}^{\phantom{\nu}\beta} + 2 \left(\nabla_{(\mu} F_{\nu)
    \beta}\right) \nabla_{\alpha} F^{\alpha \beta}  \notag \\ 
& \qquad - 2 \nabla_{\alpha} \nabla_{(\mu} \left( F_{\nu) \beta}F^{\beta
    \alpha} \right) + \Box \left(F_{\mu \beta}
  F^{\beta}_{\phantom{\beta} \nu} \right) + g_{\mu \nu}
\nabla_{\alpha} \nabla_{\beta} \left(F^{\alpha \gamma}
  F_{\gamma}^{\phantom{\gamma} \beta} \right) \notag \\ 
& \qquad - \tfrac{1}{2}\nabla_{\mu} \nabla_{\nu} F^2  + \tfrac{1}{2}
g_{\mu \nu} \Box F^2 + \tfrac{1}{2} F^2 R_{\mu \nu} - \tfrac{1}{4}
g_{\mu \nu} R F^2 - F_{\alpha \beta}F^{\beta \gamma}F_{\gamma
  \delta}F^{\delta \alpha} g_{\mu \nu} \notag \\ 
& \qquad  - F_{\alpha \beta} F^{\beta \gamma}R_{\gamma}^{\alpha} g_{\mu
  \nu} + \nabla_{\alpha}(F^{\alpha}_{\phantom{\alpha} \beta}
\nabla_{\gamma} F^{\gamma \beta}) g_{\mu \nu} - \tfrac{1}{8} F^4
g_{\mu \nu} -  \tfrac{1}{2} \left(\nabla_{\alpha}
  F^{\alpha}_{\phantom{\alpha} \beta} \right) \nabla_{\gamma}
F^{\gamma \beta} g_{\mu \nu} 
\end{align}
Here $P_{\mu \nu}$ contain all the $\mathcal{O}(\kappa)$ terms
which do not contain the field strength tensor, while $Q_{\mu
  \nu}$ terms are the $\mathcal{O}(\kappa)$ terms that do.
Variation of the action with respect to $A_{\mu}$, gives the
equation of motion
\begin{align}
\nabla_{\alpha} F^{\alpha \nu} &= 
- \kappa\left[\nabla_{\alpha}\left( F^{\alpha \nu}
\left(\frac{1}{2} R + \frac{1}{2} F^2\right)\right) 
- 4 \nabla_{\alpha}\left( F^{\alpha \beta} F_{\beta \gamma} 
F^{\gamma \nu}\right)  - \nabla_{\alpha}
\left( R^{\alpha \beta}F_{\beta}^{\phantom{\beta} \nu} 
- R^{\nu \beta}F_{\beta}^{\phantom{\beta} \alpha}\right)
\right. \nonumber \\ 
& \qquad \qquad \qquad \left. -\frac{1}{2} \Box 
\left( \nabla_{\alpha}F^{\alpha \nu}\right) + 
\frac{1}{2} \nabla_{\alpha} \nabla^{\nu}
 \left(\nabla_{\beta} F^{\beta \alpha} \right) \right]
+ {\cal O}(\kappa^2)\,.
\label{eom.maxwell}
\end{align}

Even at first order in $\kappa$\,, what we get from the
Eddington-Born-Infeld theory in five dimensions differs from what
we would get by adding usual Maxwell action to the four-dimensional
theory as in Eq.~(\ref{ebi.ebi}). This is true for all the
equations derived thence, be it Eq.~(\ref{ebi.ric2}) derived via a
Palatini variation, the expansion of that as in
Eq.~(\ref{ebi.ricq}), or Eq.~(\ref{ebi.Rexp}) derived through the
variation of a completely metric theory. The difference lies
primarily in the fact that we have couplings between the
electromagnetic field strength and the curvature. Note however that
if we set $F_{\mu\nu}=0$ in Eq.~(\ref{eom.eom}), the resulting
equation agrees with Eq.~(\ref{ebi.Rexp}).

Taking the trace of the equations brings out the difference rather
dramatically. We have seen earlier that if we add electromagnetism
as a separate matter action, the trace of Eq.~(\ref{ebi.ricq})
produces $R = 4 \Lambda$\,, since the stress-energy tensor
electromagnetism is traceless. The trace of Eq.~(\ref{ebi.Rexp})
gives the same result if we formally expand $(1 -
\tfrac{1}2{}\kappa\Box)^{-1}\,.$ In contrast, the trace of
Eq.~(\ref{eom.eom}) produces to order $\kappa$
\begin{align}
R = 4\Lambda - \kappa\left[\tfrac{1}{2}F^4 \right. &+\left.
 \tfrac{1}{2}RF^2 
+ 2 R^{\alpha\beta}F_{\beta\gamma}
F_{\phantom{\gamma}\alpha}^{\gamma}
+ 4  F^{\alpha\beta}F_{\beta\gamma}F^{\gamma\delta}
F_{\delta\alpha}\right.\nonumber \\  
 &+ \left. \nabla_{\alpha}\nabla_{\beta}\left(R^{\alpha \beta} +  
2 F^{\alpha\delta}F_{\delta}^{\phantom{\delta}\beta}\right)
- \Box\left(R - \tfrac{1}{2}F^2\right) 
+ \nabla_{\beta}F^{\beta\gamma}
\nabla^{\alpha}F_{\alpha \gamma}\right]\,. 
\label{eom.trace}
\end{align}
Thus the Kaluza-Klein prescription leads to the incorporation of 
electromagnetic fields in the theory as expected, but also to
novel non-trivial couplings which get naturally introduced 
because of the determinantal form of the action.

\section{Iterative solutions}
The equations of motion for the metric and the vector potential
will have even more complicated couplings at higher orders of
$\kappa$ as they come from a higher-order expansion of the
action~(\ref{kk.ebiaction}). However, it is possible to find
solutions to these equations to any order in $\kappa$ via an
iterative procedure, which we will now describe.

Let us rewrite the ${\cal O}(\kappa)$ equations of motion 
Eq.~(\ref{eom.eom}) and Eq.~(\ref{eom.maxwell})
as 
\begin{align}
\Lambda g_{\mu\nu} &= - G_{\mu\nu} + 8 \pi T_{\mu\nu}  
- \kappa C_{\mu\nu}\,, \\
\nabla_\mu F^{\mu\nu} &= -\kappa D^\nu\,.
\end{align}
We can split the fields $g_{\mu\nu}$ and $A_\mu$ in their zeroth
and first order parts, 
\begin{equation}
g_{\mu\nu} = g^{0}_{\mu\nu} +  g^{1}_{\mu\nu}\,, \qquad 
A_\mu = A^{0}_\mu + A^{1}_\mu\,,
\end{equation}
where $g^{0}_{\mu\nu}$ and $A^{0}_\mu$ satisfy the zeroth order
equations,
\begin{equation}
\Lambda g^{0}_{\mu\nu} = - G^{0}_{\mu\nu} +  8\pi
T^{0}_{\mu\nu} 
\,,  
\qquad \nabla^{0}_{\mu} F^{0 \mu \nu} = 0 \,,
\label{sol.rnds}
\end{equation}
with $\nabla^0_\mu\,, G^{0}_{\mu\nu}\,,$ and $  T^{0}_{\mu\nu}$
defined in terms of the zeroth order fields, and $ g^{1}_{\mu\nu}\,, 
A^{1}_\mu$ are linear in $\kappa\,.$

Let us consider the spherically symmetric case, then we have the
Reissner-Nordstr\"om-de Sitter solution for the lowest order
equations, 
\begin{equation}
g^{0}_{\mu\nu} = 
\begin{pmatrix}
- f(r)  & 0 & 0 & 0 \\ 
0 & {f(r)}^{-1} & 0 & 0\\
0 & 0 & r^2 &  0\\
0 & 0 & 0 & r^2 {\text{sin}}^{2} {\theta}\\
\end{pmatrix}\, ,
\label{met.rnds}
\end{equation}
with $f(r) = 1 - \frac{2m}{r} - \frac{\Lambda
  r^2}{3} + \frac{q^2}{r^2}\,,$ and the corresponding
\begin{equation}
F^{0}_{\mu\nu} = 
\begin{pmatrix}
0 & \frac{q}{r^2} & 0 & 0 \\ 
- \frac{q}{r^2} & 0 & 0 & 0\\
0 & 0 & 0 &  0\\
0 & 0 & 0 & 0\\
\end{pmatrix}\, .
\label{far.rnds}
\end{equation}

We can write the equations at the next order in $\kappa$ as 
\begin{align}
\Lambda g^{0}_{\mu \nu} + \Lambda g^{1}_{\mu \nu} & = -
G^{0}_{\mu\nu} - G^{1}_{\mu\nu} +  8\pi T^{0}_{\mu\nu} +
8\pi T^{1}_{\mu\nu} - \kappa C_{\mu \nu} \, , \notag\\ 
\nabla^{0}_{\mu} F^{0\mu \nu} & + \nabla^{0}_{\mu}  F^{1\mu
  \nu} + \nabla^{1}_{\mu}  F^{0\mu \nu} = - \kappa D^{\nu}
\,, 
\label{eom.kap}
\end{align}
where $G^1_{\mu\nu}$ and $T^1_{\mu\nu}$ are the ${\cal O}(\kappa)$
parts of $G_{\mu\nu}$ and $T_{\mu\nu}\,,$ and $C_{\mu \nu}$ and 
$D^{\nu}$ are defined as functions of the zeroth-order fields 
$g^0_{\mu\nu}$ and $F^0_{\mu\nu}\,,$ and have
the form
\begin{align}
C_{\mu \nu} &= \left[(S_{\mu\nu} - \frac{1}{4}g_{\mu\nu}S)  - \nabla_{\alpha}
F^{\alpha}_{\text{\space \space} \nu} \nabla_{\beta}
F^{\beta}_{\text{\space \space} \mu} - 2 F_{\mu \alpha}
(R^{\alpha \beta} + 2 F^{\alpha \gamma} F_{\gamma} ^{\text{\space
    \space} \beta})F_{\beta \nu} \right. \nonumber \\
& \left. \qquad\qquad \qquad \qquad \qquad \qquad  - H^{'}_{\mu
    \nu} - \frac{1}{2}F^2 
(R_{\mu \nu} + F_{\mu \beta} F_{\nu} ^{\phantom{\nu}\beta}) \right]\,,
\label{iterate.Cmunu} 
\\ 
D^{\nu} & =  \left[\nabla_{\alpha}\left( F^{\alpha
      \nu}\left(\frac{1}{2} R + \frac{1}{2} F^2\right)\right) - 4
  \nabla_{\alpha}\left( F^{\alpha \beta} F_{\beta \gamma} F^{\gamma
      \nu}\right)  - 2 \nabla_{\alpha}\left( R^{\alpha
      \beta}F_{\beta}^{\phantom{\beta} \nu}\right)
\right. \nonumber \\ 
& \qquad \qquad \qquad \qquad\qquad \left. -\frac{1}{2} 
\Box \left(\nabla_{\alpha}F^{\alpha \nu}\right) + \frac{1}{2}
  \nabla_{\alpha} \nabla^{\nu} \left(\nabla_{\beta} F^{\beta
      \alpha} \right) \right]\,.
\label{iterate.Dmu}
\end{align}

Given $g^0_{\mu\nu}$ and $F^0_{\mu\nu}\,,$ 
Eq.~(\ref{eom.kap})  reduces to finding the 
solution to 
\begin{align}
\Lambda g^{1}_{\mu \nu} & = - G^{1}_{\mu\nu} +
8 \pi T^{1}_{\mu\nu} - \kappa C_{\mu \nu} \, , \notag\\ 
\nabla^{0}_{\mu} F^{1\mu \nu} &+ \nabla^{1}_{\mu} F^{0\mu
  \nu} = - \kappa D^{\nu} \,.
\label{eom.kap2}
\end{align}
We can calculate $C_{\mu \nu}$ and $D^{\nu}$ in a straightforward 
manner for the Reissner-Nordstr\"om-de Sitter solution, 
\begin{align}
\kappa C_{\mu \nu} &= 
\begin{pmatrix}
- f(r) ( - \frac{9 \, \kappa \, q^4}{2 \, r^8} - \frac{\kappa  \,
  \Lambda \, q^2}{r^4})  & 0 & 0 & 0 \\  
0 & ( - \frac{9 \, \kappa \, q^4}{2 \, r^8} - \frac{\kappa \,
  \Lambda \, q^2}{r^4}) (f(r))^{-1} & 0 & 0\\ 
0 & 0 & r^2 (\frac{3 \, \kappa \, q^4}{2 \, \Lambda \, r^8} +
\frac{\kappa \, q^2}{r^4}) &  0\\ 
0 & 0 & 0 & r^2 {\text{sin}}^{2} {\theta}  (\frac{3 \, \kappa \,
  q^4}{2 \, \Lambda \, r^8} + \frac{\kappa \, q^2}{r^4})\\ 
\end{pmatrix}\, , \notag \\
& \notag \\ 
\kappa D^{\nu} &= \left(\frac{12 \kappa \, q^3}{r^7}, 0, 0,
  0\right) \,.  
\label{eom.corr2}
\end{align}
%
As might have been expected, the metric at this order remains
spherically symmetric, and that the ${\cal O}(\kappa)$ correction
for $A^{0}_{\mu}$ is only for the time component.  If we now write
the metric $g_{\mu \nu}$ and the four potential $A_{\mu}$
\begin{align}
g_{\mu \nu} &= 
\begin{pmatrix}
- f(r) - \kappa A(r)  & 0 & 0 & 0 \\ 
0 & (f(r) + \kappa A(r))^{-1} & 0 & 0\\
0 & 0 & r^2 &  0\\
0 & 0 & 0 & r^2 {\text{sin}}^{2} {\theta}\\
\end{pmatrix}\, ,  \\ 
A_{\mu} &= \left(\frac{q}{r} + \kappa B(r), 0, 0, 0\right) \, .
\label{eom.ans}
\end{align}

Putting these in Eq.~(\ref{eom.kap2}), 
we find that to $\mathcal{O}({\kappa})$, the following are the
solutions for the metric and the electromagnetic field, 
\begin{align}
g_{\mu \nu} &= 
\begin{pmatrix}
- \left(f(r) + \kappa\left(\frac{3 \, q^4}{10 \, r^6} - \frac{\Lambda \,
  q^2}{r^2} \right)\right)  & 0 & 0 & 0 \\  
0 &  \left(f(r) + \kappa\left(\frac{3 \, q^4}{10 \, r^6} - \frac{\Lambda \, 
  q^2}{r^2} \right)\right)^{-1}
& 0 & 0\\ 
0 & 0 & r^2 &  0\\
0 & 0 & 0 & r^2 {\text{sin}}^{2} {\theta}\\
\end{pmatrix}\, ,
\label{eom.fans}\\
F_{\mu \nu} &=  
\begin{pmatrix}
0  & \frac{q}{r^2} + \frac{3 \, \kappa \, q^3}{r^6} & 0 & 0 \\ 
- \frac{q}{r^2} - \frac{3 \, \kappa \, q^3}{r^6} & 0 & 0 & 0\\
0 & 0 & 0 &  0\\
0 & 0 & 0 & 0\\
\end{pmatrix}\, .
\end{align}
%
It can also be shown that the above expressions satisfy the
$\mathcal{O}(\kappa)$ trace equation 
Eq.~{\ref{eom.trace}, thus confirming that it is indeed the
solution  at this order.  The trace equation for this solution
works out to 
\begin{equation}
R = 4\Lambda + \frac{6\kappa q^4}{r^8}\,,
\end{equation}
which indicates that there is no singularity at finite $r$ for this
solution. 

It is easy to see how the iterative procedure can be extended to
include higher order terms in $\kappa$. For example, at the next
order in $\kappa$, we will have to expand the action of
Eq.~(\ref{kk.action}) to ${\cal O}(\kappa^3)$. This will produce
equations for the ${\cal O}(\kappa^2)$ terms of the metric and the
electromagnetic field tensor, which look like
\begin{align}
\Lambda g^{(2)}_{\mu \nu} & = - G^{(2)}_{\mu\nu} +
8 \pi T^{(2)}_{\mu\nu} - \kappa C'_{\mu \nu} \, , \notag\\ 
\nabla^{0}_{\mu} F^{(2)\mu \nu} &+ \nabla^{1}_{\mu} F^{1\mu
  \nu} + \nabla^{(2)}_{\mu} F^{0\mu \nu} = - \kappa D^{\prime\nu} , 
\label{eom.kap3}
\end{align}
where now $G^{(2)}_{\mu\nu}\,, T^{(2)}_{\mu\nu}$ and 
$F^{(2)\mu \nu}$ are the
$\mathcal{O}({\kappa}^2)$ terms in $G_{\mu\nu}\,, T_{\mu\nu}\,,$ 
and $F^{\mu \nu}\,,$ and $C'_{\mu \nu}$ and $D^{\prime\nu}$ 
are $\mathcal{O}({\kappa})$
functions analogous to Eq.~(\ref{iterate.Cmunu}) and
Eq.~(\ref{iterate.Dmu}), but including additional terms coming from
the higher order expansion of the action, and calculated using the
$\mathcal{O}({\kappa})$ field solutions. Extending in this manner,  we 
can find a solution to the theory to any order in $\kappa$. 

We started out with the goal of finding a natural way of adding
matter fields to the Eddington-inspired Born-Infeld action of
gravity. The Kaluza procedure is certainly natural in the sense
that it is geometric, but it deviates from the philosophy of the
Eddington action in that it has to be written purely as a metric
theory, otherwise it would not be possible to interpret $g_{\mu5}$
as components of the electromagnetic potential.

Since the five-dimensional action comes from the expansion of the
square root of a polynomial, it is non-polynomial in nature, and
thus the coupling between electromagnetism and gravity is highly
nonlinear.  However, we can expand in powers of curvature
(equivalently in powers of $\kappa$) and find and solve the
equations of motion term by term. Applying this iterative procedure
to the lowest order electric Reissner-Nordstr\"om-de Sitter black
hole solution, we found the ${\cal O}(\kappa)$ correction. At least
at this order, there is no singularity at finite radius, unlike the
`surface singularities' which plague four-dimensional
Eddington-inspired Born-Infeld gravity with minimally added matter.

\begin{acknowledgements}
We thank the anonymous referee, whose critical comments helped us 
make significant additions to the paper. 
\end{acknowledgements}

\appendix

\section{Non-trivial $\Phi$}\label{appb}

If we do not set $\Phi=1$ in the five-dimensional metric 
Eq.~(\ref{kk.ansatz}), we will find the following Ricci tensor components,
\begin{align}
\hat{R}_{\mu\nu} &= R_{\mu\nu} - \Phi^{-1}{\nabla}_{\mu}
{\nabla}_{\nu} 
  {\Phi} + \frac{{\alpha}^4}{4} F^{\beta\gamma}F_{\beta\gamma}
  {\Phi}^4 A_{\mu}A_{\nu} 
 - \frac{{\alpha}^2}{2} \left[3
{\Phi}{\partial}_{\beta}{\Phi}\left(F^{\beta}_{\phantom{\beta}
  \mu}A_{\nu} + F^{\beta}_{\phantom{\beta}\nu}A_{\mu}\right) \right.
\notag \\ 
&\qquad\left. + 2A_{\mu}A_{\nu}{\Phi} \Box {\Phi} + {\Phi}^2
\left(A_{\mu}{\nabla}_{\beta} F^{\beta}_{\phantom{\beta} \nu} +
A_{\nu}{\nabla}_{\beta} F^{\beta}_{\phantom{\beta} \mu} +
F_{\beta\mu}F^{\beta}_{\phantom{\beta}\nu}\right)\right]
\\
\hat{R}_{\mu 5} &= \frac{{\alpha}^3}{4} F^{\beta \gamma}F_{\beta
  \gamma} {\Phi}^4 A_{\mu} - 
\frac{\alpha}{2} \left(3{\Phi}{\partial}^{\beta}{\Phi}F_{\beta \mu}
  + {\Phi}^2{\nabla}_{\beta}F^{\beta}_{\phantom{\beta}\mu} 
+ 2{\Phi} \Box {\Phi}A_{\mu}\right) 
\\
\hat{R}_{55} &= \frac{{\alpha}^2}{4} {\Phi}^4 F^{\beta \gamma}
F_{\beta \gamma} - {\Phi} \Box {\Phi}\,.
\end{align} 

The five-dimensional Ricci scalar is calculated from this to be
\begin{equation} 
\hat{R} = R - 2 {\Phi}^{-1} \Box {\Phi} - \frac{{\alpha}^2}{4}
{\Phi}^2 F^{\beta \gamma} F_{\beta \gamma}\,.
\end{equation}
%



\end{document}